\documentstyle[pra,aps]{revtex}
\begin{document}
\draft
\author{V.B. Svetovoy and M.V. Lokhanin}
\address{{\small Department of Physics, Yaroslavl State University,} \\ {\small %
Sovetskaya 14, Yaroslavl 150000, Russia}}
\title{Comment on the temperature dependence of the Casimir force.}
\maketitle

\begin{abstract}
Linear in temperature correction to the Casimir force is discussed. The
correction is important for small separations between bodies tested in the
recent experiments and disappears in the case of perfect conductors.
\end{abstract}


The Casimir force \cite{Casimir} has been measured with high precision in
recent experiments \cite{Lam1,MR,RLM}. Also there are plans \cite{Long,Fisch}
to look for very weak hypotetical forces where the Casimir force is the main
background. All this makes the precise evaluation of the Casimir force an
important problem. Here we will discuss a particular problem concerning the
temperature dependence of the force between macroscopic bodies made of
nonideal metals.

For perfect conductors the temperature correction has been found many years
ago \cite{Meh,Brown,Schw} and it is small for small separations between
bodies $a\ll c\hbar /kT$ or equivalently for low temperature. For a sphere
above a disk the leading term behaves as $\left( T/T_{eff}\right) ^3$, where 
$T_{eff}=\hbar c/2a$. This result follows from a general expression for the
Casimir force given by Lifshitz \cite{Lif,LP} modified for the case of
sphere-disk geometry with the proximity force theorem \cite{PFT}:

\begin{equation}
\label{shpl}F(a)=-\frac{kTR}{c^2}{\sum\limits_{n=0}^\infty {}}^{\prime
}\zeta _n^2\int\limits_1^\infty dpp\ln \left[ \left( 1-G_1e^{-2p\zeta
_na/c}\right) \left( 1-G_2e^{-2p\zeta _na/c}\right) \right] , 
\end{equation}

\noindent where $R$ is the sphere radius,

$$
G_1=\left( \frac{p-s}{p+s}\right) ^2,\quad G_2=\left( \frac{\varepsilon
\left( i\zeta _n\right) p-s}{\varepsilon \left( i\zeta _n\right) p+s}\right)
^2,\quad 
$$

\begin{equation}
\label{defin1}s=\sqrt{\varepsilon \left( i\zeta _n\right) -1+p^2},\quad
\zeta _n=\frac{2\pi nkT}\hbar , 
\end{equation}

\noindent $\varepsilon \left( i\zeta _n\right) $ is the dielectric function
of the used material at imaginary frequencies. The prime over the sum sign
indicates that the first term $n=0$ has to be taken with the coefficient $%
1/2 $.

For small temperature the sum in (\ref{shpl}) can be replaced by the
integral over $\zeta $ and the resulting force does not depend on the
temperature at all. In general, the replacement is true with the precision $%
\sim T/T_{eff}$. In condition of the atomic force microscope experiments 
\cite{MR,RLM} the smallest separation was $0.1\ \mu m$ and the replacement
error can be as large as 3\%. It exceeds the experimental errors $\sim $1\%
and, therefore, the finite temperature effect has to be taken into account.
We define the temperature correction $\Delta _TF$ as difference between
forces written as the sum over $n$ and as the integral instead of this sum.

Special care needs to treat the first term $n=0$ in Eq.(\ref{shpl}). The
formal reason is that $\zeta _n^2$ becomes zero but the integral over $p$
diverges. The physical reason is that this term corresponds to the static
limit when for metallic bodies $\varepsilon \rightarrow \infty $. This means
that any parameter characterizing the dielectric function of a metal cannot
appear in the $n=0$ term in contrast with a dielectric for which it will
depend on the static permittivity of the material. In the $\varepsilon
\rightarrow \infty $ limit the functions $G_{1,2}$ become $G_1=G_2=1$. The
formal problem is overcome by introducing the integration over a new
variable $x=2p\zeta _na/c$ and after that one can take $\zeta _n=0$ for the $%
n=0$ term. Transformed in this way Eq.(\ref{shpl}) will be

\begin{equation}
\label{base}F(a)=\frac{kTR}{4a^2}\left\{ \zeta \left( 3\right) -{%
\sum\limits_{n=1}^\infty {}}\int\limits_{x_n}^\infty dxx\ln \left[ \left(
1-G_1e^{-x}\right) \left( 1-G_2e^{-x}\right) \right] \right\} , 
\end{equation}

\noindent where $\zeta \left( m\right) $ is the zeta-function and

\begin{equation}
\label{xn}x_n=\frac{2\zeta _na}c. 
\end{equation}

\noindent Here the first term is linear in temperature and it corresponds to
the $n=0$ term in (\ref{shpl}).

The sum in (\ref{base}) as a function of temperature contains a piece linear
in $T$ which exactly cancels for ideal metals the first term giving the well
known result

\begin{equation}
\label{FT}F_T(a)=F_0(a)\left[ 1+\frac{45\zeta \left( 3\right) }{\pi ^3}%
\left( \frac T{T_{eff}}\right) ^3-\left( \frac T{T_{eff}}\right) ^4\right] , 
\end{equation}

\noindent where $F_0(a)=\pi ^3\hbar cR/(360a^3)$ is the bare Casimir force
between sphere and plate. (\ref{FT}) is written in the small temperature
limit when corrections to $F_0(a)$ are very small.

If we are using the dielectric function of a real metal, the cancellation of
the first term in (\ref{base}) can be incomplete and the linear in $T$
contribution can survive. That was noted first in \cite{SL}, where Eq.(\ref
{base}) was used for numerical calculation of the Casimir force. It was
found that for the experiments \cite{MR,RLM} the temperature correction at
the smallest separation is $4\ pN$ against the experimental errors $2\ pN$.
This conclusion has been criticized in Ref.\cite{BGKM}, where the linear
correction was not found. In this connection we would like to clarify here
difference in the approaches.

The $n=0$ term was discussed in \cite{BGKM} on the right basis but for
actual calculations the following expression has been used

\begin{equation}
\label{alt}F(a)=-\frac{kTR}{4a^2}{\sum\limits_{n=0}^\infty {}}%
^{\prime}\int\limits_{x_n}^\infty dxx\ln \left[ \left( 1-G_1e^{-x}\right) \left(
1-G_2e^{-x}\right) \right] , 
\end{equation}

\noindent where for $n=0$ the function $G_1\neq 1$. It is clear from the
expression for the force in the high temperature limit, where only the $n=0$
term survives (Eq.(16) in \cite{BGKM})

\begin{equation}
\label{hT}F(a)=\frac{kT}{4a^2}R\zeta \left( 3\right) \left( 1-\frac{2c}{%
a\omega _p}\right) . 
\end{equation}

\noindent Here $\omega _p$ is the plasma frequency of the used metal. The
parameter $\omega _p$ in this equation shows that the special prescription
for the $n=0$ term has not been done. The dielectric function was described
by the plasma model where it is

\begin{equation}
\label{plasma}\varepsilon \left( i\zeta \right) =1+\frac{\omega _p^2}{\zeta
^2}. 
\end{equation}

\noindent In the high temperature limit only low frequency fluctuations are
important and in this range metals can be much better described by the Drude
dielectric function

\begin{equation}
\label{Drude}\varepsilon \left( i\zeta \right) =1+\frac{\omega _p^2}{\zeta
\left( \zeta +\omega _\tau \right) }, 
\end{equation}

\noindent where $\omega _\tau $ is the relaxation frequency. However, if we
use (\ref{Drude}) to find the classical limit with the help of (\ref{alt}),
the result will be wrong, namely, two times smaller than the well known
limit $kTR\zeta \left( 3\right) /4a^2$. Eq.(\ref{base}) does not suffer from
this problem.

The authors \cite{BGKM} convincingly demonstrated that for low temperatures
Eq.(\ref{alt}) does not give the correction linear in $T$ and the leading
correction is only $\left( T/T_{eff}\right) ^3$. One can use this result to
extract the linear term from the sum in Eq.(\ref{base}) explicitly. The
difference between (\ref{base}) and (\ref{alt}) gives the correction we are
looking for if one neglects the higher order terms in $T/T_{eff}$

\begin{equation}
\label{delT}\Delta _TF=\frac{kTR}{4a^2}\left\{ \zeta \left( 3\right) +\frac 1%
2{}\int\limits_0^\infty dxx\ln \left[ \left( 1-G_1e^{-x}\right) \left(
1-G_2e^{-x}\right) \right] \right\} . 
\end{equation}

\noindent The integral here is the linear term contained in the sum in (\ref
{base}) and, of course, it can depend on the material parameters because the
summation is going over nonzero frequencies $\zeta _n$. On the other hand,
since this integral appeared as the $n=0$ term in (\ref{alt}), we should
take the functions $G_{1,2}$ at $x_n=0$. In this limit $G_2=1$ but $G_1\neq
1 $. Using then the relation $\int_0^\infty dxx\ln \left( 1-e^{-x}\right)
=-\zeta \left( 3\right) $ one finds the final expression for the correction
linear in $T$:

\begin{equation}
\label{delTf}\Delta _TF=\frac{kTR}{8a^2}\left[ \zeta \left( 3\right)
+\int\limits_0^\infty dxx\ln \left( 1-G_1e^{-x}\right) \right] , 
\end{equation}

\noindent where

$$
G_1=\left( \frac{x-\sqrt{x^2+\alpha ^{-2}}}{x+\sqrt{x^2+\alpha ^{-2}}}%
\right) ^2,\qquad \alpha =\frac c{2a\omega _p}. 
$$

\noindent Let us stress that (\ref{delTf}) is true only for the plasma
model. When $\omega _p\rightarrow \infty $ the correction disappears as it
should be. Expansion in powers of $\alpha $ gives

\begin{equation}
\label{exp}\Delta _TF=\frac{kTR}{8a^2}\zeta \left( 3\right) \cdot 8\alpha
\left( 1-3\alpha +O\left( \alpha ^2\right) \right) . 
\end{equation}

\noindent For $\omega _p=2\cdot 10^{16}\ s^{-1}$ and $a=0.1\ \mu m$ one gets 
$\Delta _TF\approx 2.5\ pN$ using (\ref{delTf}) or calculating directly with
the help of (\ref{base}) and $2.9\ pN$ using (\ref{exp}). The correction
increases further if we will use the Drude dielectric function (\ref{Drude}%
). In this case it has to be evaluated numerically using (\ref{base}) and (%
\ref{shpl}) with the integral instead of the sum. The relaxation frequency $%
\omega _\tau $ influences mostly on the integral since it changes low
frequency behavior of the itegrand. For typical value $\omega _\tau =5\cdot
10^{13}\ s^{-1}$ we found $\Delta _TF\approx 4.0\ pN$. It cannot be compared
directly with the value given in \cite{SL} because layered body cover has
been considered there but it is clear that the calculations here give the
same order of magnitude for $\Delta _TF$.

In conclusion, we have considered the linear in temperature correction to
the Casimir force at low temperatures or equivalently at small separations.
Special care has to be taken to get the contribution of the fluctuations in
the static limit ($n=0\ $term). This contribution is canceled for ideal
mirrors but cancellation is incomplete for real metals. The right treatment
of the $n=0$ term allowed to use the Drude dielectric function for metals
which is more appropriate at low frequencies than the function in the plasma
model.

\end{document}